# The influence of cold wall thickness and background activity on recovery coefficients in PET and SPECT

Robin de Nijs[1]

1. Department of Clinical Physiology and Nuclear Medicine, Copenhagen University Hospital – Rigshospitalet, Copenhagen, Denmark

## TECHNICAL NOTE

Running title: Cold wall thickness and recovery coefficients

Corresponding author
Robin de Nijs
Department of Clinical Physiology and Nuclear Medicine
Copenhagen University Hospital – Rigshospitalet
Blegdamvej 9
DK-2100 Copenhagen
Denmark
e-mail address: robin.de.nijs@regionh.dk

**Highlights**

- Novel equation for the recovery coefficient (RC) for spheres with cold walls
- RC-equation for large objects with cold walls and background activity
- General RC-equation incorporating cold wall thickness and background activity
- Cold wall effect scales with the object-to-background concentration ratio (OBR).
- For a 10 mm sphere and a 1 mm cold wall RC can change up to 0.19/OBR.

**Abstract**

*Purpose:* The partial volume effect caused by limited spatial resolution consists of signal spill-out from and, in the case of background, spill-in into objects. This can be described by the recovery coefficient (RC), which can be measured in phantoms. With background activity present the RC of an object inside a phantom is influenced by its confining cold wall. This is relevant for the use of RC-values in quantitative imaging, for standard uptake values, and dosimetry.

*Methods:* A theoretical equation for RC as a function of object-to-background concentration ratio (OBR) is formulated and combined with a correction for wall thickness. Novel analytical expressions for spheres and large objects of arbitrary size are derived for the RC of the mean ($RC_{mean}$). For spheres this expression was compared to previously published experimental data exploring thick (1.5 mm) and thin (0.18 mm) walls.

*Results:* For 10, 37 and 60 mm spheres, 1 mm wall thickness and isotropic spatially invariant Gaussian point spread function with a full width half maximum in the range 2-20 mm, the minimum-maximum values for the change in $RC_{mean}$ times OBR are 0.03-0.19, 0.05-0.06, 0.03-0.04, respectively. Experimental and theoretical differences in $RC_{mean}$ between spheres with thick and thin walls were in good agreement for 10, 15, 38 and 58 mm spheres, but not for the 20 and 30 mm spheres.

*Conclusion:* When using phantom measurements for determining RC-values it is important to consider the cold wall thickness, otherwise significant errors can arise in the presence of background activity.

## 1. Introduction

The so-called partial volume effect (PVE) is caused by the limited spatial resolution of an imaging system such as Positron Emission Tomography (PET) and Single Photon Emission Computed Tomography (SPECT). The discretization into voxels can also contribute to this effect[1]. Spatial resolution can be described by a point spread function (PSF), which in some cases can be assumed to be spatially invariant and isotropic. For objects filled with activity the PVE causes a spill-out of the signal originating from the object, while the surrounding activity, often the background, causes a spill-in of signal into the object.

### *1.1 Definitions of recovery coefficients*

One method for correcting the PVE is the application of recovery coefficients (RC). In phantom studies both the background and objects inside the phantom often contain a uniform activity distribution or no activity. For objects the RC and contrast recovery coefficient (CRC)[2,3] are defined by

$$RC = \frac{c}{a} \quad \text{and} \quad CRC = \frac{c-c_{\mathrm{BG}}}{c_{\mathrm{BG}}} \Big/ \frac{a-a_{\mathrm{BG}}}{a_{\mathrm{BG}}} = \frac{OBR^{*}-1}{OBR-1} = \frac{c-c_{\mathrm{BG}}}{a-a_{\mathrm{BG}}} \cdot \frac{1}{RC_{\mathrm{BG}}}, \qquad (1)$$

where $a$ refers to the true activity concentration in the object and $a_{\mathrm{BG}}$ to the true background concentration. Measured object and background concentrations are denoted by $c$ and $c_{\mathrm{BG}}$, respectively. Background concentration is measured far away from the object, so that spill-out from the object is negligible. Alternatively, the CRC is defined by means of the true object-to-background activity concentration ratio ($OBR = a/a_{\mathrm{BG}}$) and measured object-to-background activity concentration ratio ($OBR^{*} = c/c_{\mathrm{BG}} = OBR \cdot RC/RC_{BG}$). The recovery coefficient for the background is denoted by $RC_{\mathrm{BG}}$. This is an extra term compared to the definition of Kessler et al[2] and complies with the NEMA- and EARL-definitions[3–5]. For a well-calibrated scanner $RC_{\mathrm{BG}}$ equals unity and this is also case if the background is used for the activity calibration. In the case of zero-background there is no spill-in and the “hot spot” RC[2] is equal to CRC for $RC_{\mathrm{BG}}$=1. Some definitions of CRC as the first two in equation (1) [3] and the NEMA-definition[4,5] result in division by zero in the absence of background activity. This is avoided by the introduction of $RC_{\mathrm{BG}}$, which is called “calibration bias” in Sunderland et al[3], where OBR is mentioned as “contrast (ratio)”. Since CRC equals RC for zero background for $RC_{\mathrm{BG}}$=1, it is at least theoretically independent of the background activity concentration[2]. CRC as defined in equation (1) is not influenced by bias of the scanner calibration, since both measured and true concentration differences are expressed relatively to their respective background concentration.

*1.2 Spill-out, spill-in and the effect of background activity*

In figure (1) the partial volume effect, spill-in and spill-out are illustrated for a disc. RC- and CRC-values can be calculated for both the mean and maximum value of the activity concentration inside the object. In this paper the main focus is on the (C)RC of the mean. By combining the definitions for RC and CRC in equation (1) the recovery coefficient *RC* can be described as

$$RC = RC_{\mathrm{out}} + RC_{\mathrm{in}} = RC_{BG} \cdot \{CRC + OBR^{-1} \cdot (1 - CRC)\} \tag{2}$$

where the first term $RC_{out}$ equals RC only taking spill-out into account, while the second term $RC_{in}$ adjusts RC for spill-in from the background activity[6,7]. For spheres and objects of arbitrary shape the so-called sphere-to-background ratio SBR and the target-to-background ratio TBR are often used as an expression for OBR. Note that equation (2) also is valid for the case of zero-background ($OBR^{-1} = 0$), if a fixed value for $RC_{\mathrm{BG}}$ is assumed. This is sensible, since $RC_{\mathrm{BG}}$ is conceptually similar to the definition of the calibration factor for converting count rate concentration to activity concentration. Especially, in advanced iterative reconstruction methods that include resolution modeling non-linear volume- and contrast-dependent behavior, and convergence issues have been observed[8,9]. Therefore, caution is needed when using equation (2), since the implicit assumptions that the background RC (calibration bias) is constant when varying OBR, the terms for spill-out and spill-in can be added and CRC (and PSF) for the spill-in and spill-out term are identical and not depending on OBR might be violated. However, equation (2) holds for a "linearly additive system" [2].

### *1.3 Objects in phantoms and the cold wall effect*

An object inside a phantom with a confining wall often has a non-negligible wall thickness as is the case for the hollow spheres inside the NEMA (North-American Electrical Manufacturers Association) Body phantom[4,5], where the spheres with a diameter of 10, 13, 17, 22, 28 and 37 mm may have a wall thickness up to 1 mm. The wall does not contain activity and will therefore reduce the spill-in and decrease the RC, when background activity is present, while spill-out is not affected. In the Methods section a modified version of equation (2) for taking cold walls into account will be derived. For spheres it will be applied for the calculation of RC for the mean activity concentration. This is for instance relevant in dosimetry for the calculation of average organ dose, standard uptake values and calculation of the spatial resolution based on RC-curves[6]. With the resulting equations it is possible to calculate the RC for the mean (and maximum) for spheres based on background activity concentration, wall thickness, diameter and spatial resolution. It will be shown that a cold wall can have a significant effect on RC-values. There are a limited number of experimental studies of the cold wall effect. Hofheinz et al[10] studied the influence of a cold wall on the volume reproducing threshold, but not recovery coefficients. The latter was done by Berthon et al[11], who studied the difference for recovery coefficients (and other parameters) between spheres with different wall thicknesses, but with same inner volume. Both studies showed that a significant of the cold wall can arise.

## 2. Methods

### *2.1 A modified equation for RC incorporating cold wall thickness*

For calculating the RC for an object with a cold wall equation (2) needs to be modified. In the following it is assumed that the background is used for activity calibration, i.e. measured and true background concentration are equal and $RC_{BG}$=1. It is also assumed that CRC is independent of the background concentration. Since the cold wall does not affect the spill-out, only the term for the spill-in needs to be modified. By replacing the CRC with $CRC_t$ equation (2) becomes

$$RC_t = CRC + OBR^{-1} \cdot (1 - CRC_t) = RC - OBR^{-1} \cdot \delta CRC \quad \text{with} \quad \delta CRC = CRC_t - CRC, \quad (3)$$

where the term $OBR^{-1} \cdot \delta CRC$ equals the change in the recovery coefficient ΔRC due to the combination of background activity and wall thickness. The spill-in term $(1 - CRC_t)$ shows that $CRC_t$ can be calculated by reversing the true activity concentrations. This means that the true activity concentration inside the object including the wall is set to the original true background concentration and that the true background concentration is set to zero. $CRC$ equals $CRC_t$ for $t$=0 and $CRC_t$ is larger than CRC for non-zero wall thicknesses. Equation (2) and (3) for recovery coefficients are similar to equations (5) and (8) for activity (concentration) profiles in Hofheinz et al[10]. RC-values for the maximum can directly be calculated using equation (3) utilizing the point value of CRC for the object without wall (spill-out) and the $CRC_t$-value for the object with wall (spill-in). For a sphere a theoretical equation for the maximum RC-value is known, if a spatially invariant isotropic Gaussian PSF is assumed[2,7].

*2.2 Contrast recovery coefficient for large objects*

For the calculation of $CRC_t$ for the mean the activity concentration needs to be integrated over the object without the wall yielding the apparent amount of activity inside the inner volume, which divided by the true activity amount yields the RC. For large objects with arbitrary shape the theoretical contrast recovery coefficient[7] is given by

$$CRC_t = 1 - \frac{S}{V}\delta_t \quad \text{with} \quad \delta_t = \frac{\sigma}{\sqrt{2\pi}} e^{-\frac{t^2}{2\sigma^2}} - \frac{t}{2}\left[1 - \text{erf}\left(\frac{t}{\sigma\sqrt{2}}\right)\right] \approx \frac{\sigma}{\sqrt{2\pi}} - \frac{t}{2}, \tag{4}$$

where $S$ is the surface area, $V$ the volume, $\delta_t$ the smoothing thickness for an object with a cold wall and $\sigma$ the spatial resolution from the isotropic spatially invariant Gaussian PSF, which also can be expressed as full width half maximum ($\text{FWHM} = \sigma \cdot \sqrt{8\ln 2}$). The right-hand side of equation (4) is the first order approximation in $t$. The smoothing thickness is calculated in the same way as equation (7) in de Nijs[7], where a one-dimensional (1D) Heaviside step function $H(x)$ is convolved with a 1D Gaussian PSF, followed by integration over the intensity profile from -∞ to 0. In the case of an object with a cold wall and reversed true activity concentrations the step position x=0 corresponds by choice to the outer wall position of the object. The signal loss from $x=t$ (the inner wall position) to ∞ is equal to integration of the intensity profile from -∞ to $-t$. The smoothing thickness $\delta_t$ in equation (4) is now directly found by using these latter integration limits for $x$ in equation (II.3) in de Nijs[7]. The change in contrast recovery coefficient due to the cold wall is consequently in the first order approximation given by

$$\delta CRC = CRC_t - CRC \approx \frac{t}{2}\frac{S}{V} = \frac{3t}{2r}, \qquad (5)$$

where the right-hand side equals the change for a large sphere with radius $r$, or the so-called generalized radius $\rho = 3V/S$ can be applied for other shapes. In this approximation the change in RC due to finite wall thickness is not influenced by the spatial resolution of the scanner and it can be considered as an upper bound approximation. The latter equation can also be understood by assuming half the original background activity concentration (for a blurred measurement) uniformly distributed in the wall and calculating the signal loss in the concentrical shell.

In the NEMA-standards[4,5] a wall thickness up to 1 mm is allowed. Assuming this (typical) wall thickness of 1 mm and a 10 mm diameter sphere the change according to equation (5) is 0.3. For calculation of ΔRC this value needs to be divided by SBR, which typically is between 2:1 and 10:1 in phantom measurements. Sphere to background ratios of 4:1 (NEMA NU-2 2018 [5] for PET, section 7), 8:1 (NEMA NU-1 2023[4] for SPECT, section 6) and 10:1 (EARL[3,12] PET/CT accreditation) are examples of used ratios, so the influence of wall thickness can be significant.

For quantitative imaging, dosimetry and determination of spatial resolution the RC is measured as a function of object size and typically plotted as a function of the radius or its reciprocal, since the RC for large objects is inversely proportional to the generalized radius. For plotting the wall thickness the ratio between $t$ and $\sigma$ is the relevant parameter. In practice the scanner's spatial resolution is typically significantly smaller than the object size and the wall thickness is significantly smaller than the scanner's spatial resolution.

### *2.3 Contrast recovery coefficient for a sphere with a cold wall*

In order to calculate the influence of a finite wall for a sphere the original equation for the spherical recovery coefficients as derived by Gabiña et al[13] is modified. The wall thickness dependent contrast recovery coefficient in a sphere is given by

$$CRC_t = \mathrm{erf}\left(\frac{2r+t}{\sigma\sqrt{2}}\right) + \frac{3r^2t+3rt^2+t^3}{2r^3}\left[\mathrm{erf}\left(\frac{2r+t}{\sigma\sqrt{2}}\right) - \mathrm{erf}\left(\frac{t}{\sigma\sqrt{2}}\right)\right] + \frac{\sigma}{\sqrt{2\pi}}\left[\frac{r^2+rt+t^2-\sigma^2}{r^3}\mathrm{e}^{-\frac{(2r+t)^2}{2\sigma^2}} - \frac{3r^2+3rt+t^2-\sigma^2}{r^3}\mathrm{e}^{-\frac{t^2}{2\sigma^2}}\right]. \quad (6)$$

This equation is derived, see Appendix I, in a similar way as in Gabiña et al[13], where in the case of a cold wall and reversed true activity concentrations the limits for the integration over the activity concentration are set to the inner sphere, i.e. between 0 and *r*, while the activity is distributed over the sphere including the wall with radius *r*+*t*, In figure 2 the spill-in part of equation (3) is shown for a disc with a cold wall. The disc with reversed concentrations for calculation of $CRC_t$ is also shown. Figure 3 shows the dependence of $\delta CRC$ on diameter and wall thickness in a two-dimensional image, where the intensity scale shows the change in RC-value expressed as the background independent $\delta CRC$. All calculations are performed in Matlab R2021a (Natick, MA, USA). Equation (6) is only valid, if the imaging system can be described by for an invariant isotropic three-dimensional Gaussian PSF-function.

Equation (6) is rather difficult to comprehend and therefore some simpler approximations are derived and investigated. For large spheres $\delta CRC$ can be approximated in first order in *t* from equation (6) by

$$\delta CRC_1 = CRC_t - CRC \approx \frac{t}{r}\left(\frac{3}{2} - \frac{3}{\sqrt{2\pi}}\frac{\sigma}{r}\right), \tag{7}$$

which is consistent with equation (5). For a given $t/\sigma$-ratio the maximum $\delta CRC$ according to equation (7) can be calculated with

$$\delta CRC_{\max} \approx \frac{3}{8}\sqrt{\frac{\pi}{2}} \cdot \frac{t}{\sigma} = 0.4700 \cdot \frac{t}{\sigma} \quad \text{for} \quad \frac{\sigma}{r} = \sqrt{\frac{\pi}{8}} = 0.6267\ . \tag{8}$$

The position of the maximum is also the maximum of the $\sigma/r$ ratio up to which the approximation can be used and corresponds to corresponding to FWHM/$d$=0.74. In figure (4) an example of equation (7) and (8) is shown compared to the theoretical model of equation (3) and (6).

*2.4 Experimental difference between thick and thin walls*

The combination of equation (3) and (6) can be compared with previous experimental results as in Berthon et al[11]. In this article spheres with thin (~0.18 mm) and thick (~1.5 mm) walls are compared for several measures including RC. Differences in RC ΔRC are shown in figure 6 in Berthon et al[11] for six different TBR (=SBR) values (1.4, 2.1, 2.7, 2.8, 4.8 and 6.4) and for six different spheres with a diameter of 10, 15, 20, 30, 38 and 58 mm. By multiplying the measured RC-differences with SBR a background independent RC-difference $\Delta CRC = SBR \cdot \Delta RC$ between spheres with thin and thick walls is obtained, which corresponds to the difference for a 1:1 sphere to background ratio.

## 3. Results

*3.1 Analytical calculations for the NEMA body phantom*

Inserting equation (6) into (3) yields an equation for change in the recovery coefficient of the mean for spheres caused by a cold wall in the presence of background activity. In table 1 some values for the smallest ($d$=10 mm) and largest sphere ($d$=37 mm) of the NEMA body phantom[4,5] are shown for a wall thickness of 1 mm and different values for the scanner resolution described by the FWHM of the assumed spatially invariant isotropic Gaussian PSF. The suggested 60 mm sphere for EARL accreditation of $^{177}$Lu SPECT/CT[14] is also shown in table 1. Changes in the recovery coefficient are expressed as $\delta CRC$, which need to be divided SBR for the change in RC for a given SBR. The approximations $\delta CRC_{1st\ term}$, $\delta CRC_1$, $\delta CRC_{max}$ defined by equation (5), (7) and (8) are also included in the table.

*3.2 Absolute and relative changes in RC*

For estimations of the activity and dosimetry calculations the measured activity (concentration) is divided by the recovery coefficient and therefore the relative change in RC is relevant. Since RC also depends on SBR, it is necessary to calculate the relative difference for each SBR-value. In figure 5 both the absolute change expressed as SBR-independent $\delta CRC$ and the relative change $\Delta RC/RC$ for SBR=4 are shown for the same seven spheres as in table 1. A realistic wall thickness of 1 mm is applied, resulting in graphs that only depend on the FWHM of the imaging system. Again, the values for $\delta CRC$ give the absolute change in RC when divided by SBR.

*3.3 Comparison with experimental results*

The difference in $RC_{mean}$ between spheres with thick and thin walls is analyzed as described in section 2.4. The measured data is quite noisy (the difference in RC does not decrease monotonously), so for each sphere the six background corrected values are averaged and the standard deviation for the mean was calculated. The result is shown in figure 6 and compared to the theoretical model based on equation (3) and (6), which was used to calculate theoretical values for $\Delta CRC$. For the model a FWHM of 6 mm was assumed and wall diameters were set to averaged values of 1.530 and 0.183 mm, respectively. The CRC-difference ΔCRC is similar to the definition of δCRC in equation (3), but in this case it is effectively the difference between the CRC for a thick and thin wall instead of a difference between CRC with and CRC without a wall.

## 4. Discussion

### *4.1 Analytical calculations for the NEMA body phantom*

Inspecting table 1 shows that the change in RC of the mean can be significant, especially for small spheres. For instance, for the smallest sphere in the NEMA body phantom with a diameter of 10 mm and realistic spatial resolution for PET with FWHM=4 mm δCRC equals 0.19 meaning that the difference in $RC_{mean}$ for SBR=4 (as in NEMA NU-2[5]) is approximately 0.05. For the largest spheres of 37 and 60 mm δRC is typically 0.05-0.06 and 0.03-0.04, which makes the RC-difference of the order of 0.01 for SBR=4 indicating that the cold wall effect is rather limited. The three approximations in table 1 are shown to valid, except for the case where FWHM/d is 0.8 or larger as expected by the theory. The approximation from equation (5) $CRC_{1st\ term}$ is a very useful simple approximation indicating an upper bound for the RC-difference. Optionally also the approximation of equation (8) $CRC_{max}$ can be utilized and the lowest of these two approximations can be employed as an estimation of the upper bound of δCRC. If the upper bound value is acceptably low the cold wall effect can be neglected. For values of FWHM of 10 mm or more, as for SPECT, the cold wall effect is less pronounced, but can still be relevant for the smallest 10 mm sphere and can be of the order of 0.1 for FWHM-values of approximately 10 mm. However, for SBR=8 as in NEMA NU-1[4] the difference in RC will be of the order 0.01 which might be rather limited, when compared to the uncertainty in measured RC-values.

*4.2 Absolute and relative changes in RC*

Another way of looking at the influence of cold walls on RC-values is shown in figure (5). The left panel corresponds to the values in table 1 (again for a fixed wall thickness of 1 mm), where 0.19 is the maximum value for δRC for a FWHM-value just below 4 mm. The right panel of figure (5) show the relative changes in RC for SBR=4. Relative changes are relevant if the recovery coefficient is used for the calculation of activity (concentration) as e.g. in dosimetry and calculation of standard uptake values, where the measured value is divided by the RC-value. For the smallest sphere the curve shows changes up to 8% for a FWHM-value of approximately 6 mm. For SBR=8 (not shown) the maximum value is approximately 5% for a FWHM-value of approximately 8 mm. This shows (again) that the cold wall effect is most important for small objects and can cause significant changes in $RC_{mean}$ and thus estimated activity concentrations and calculated doses.

*4.3 Experimental difference between thick and thin walls*

The theoretically calculated values for the RC-differences in Berthon et al[11] fit very well for the smallest two and largest two spheres, but not for the spheres with a diameter of 20 and 30 mm as can be seen in figure (6). This could possibly be related to non-linear behavior of the reconstruction algorithm with resolution modelling or limitations of the model but could also be caused by the segmentation procedure. As mentioned in the introduction caution is needed in the case of reconstruction with resolution modelling and not only regarding equation (2). Equation (6) is derived for a spatially invariant three-dimensional Gaussian PSF, which is probably not a good description of the imaging system, if the reconstruction includes resolution modelling[8,9].

### *4.4 Relevance of the results and cold wall effect*

The focus in this paper is the mean recovery coefficient, since it is most important for the (mean) activity concentration estimation in phantoms. For spheres a novel equation for $CRC_t$ is derived. For the maximum recovery coefficients of spheres already known models for $CRC_t$ can be used, since it is not necessary to integrate over the inner sphere. It is enough to replace $CRC_t$ with the CRC-value for the outer sphere. Equation (3) can be used for the calculation of both the recovery coefficient for the maximum and mean. Since it is a general equation, it can be used for objects of arbitrary shape. For large objects of arbitrary shape with a cold confining wall equation (4) can be applied.

Apart from the application of the in this paper described theory to phantoms, it might be possible to use equation (3) in the calculation of image derived input functions of the aorta[15], where the aorta-wall can be described by a cold wall of a few mm and the surrounding tissue as background.

## 5. Conclusions

Cold walls in phantom experiments might cause a significant change in calculated recovery coefficients. For a linearly additive system the cold wall can be taken into account by modifying the contribution of the spill-in to the recovery coefficient. This methodology can be applied to both maximum and mean recovery coefficients. For linearly additive systems with a spatially invariant

Gaussian PSF theoretical equations for both mean and maximum recovery coefficient can be derived for objects with a cold wall as is shown for spheres. The cold wall effect typically gets larger for smaller objects, thicker walls and better spatial resolution.

## Declarations of interest

None

## Declaration of generative AI and AI-assisted technologies in the manuscript preparation process.

During the preparation of this work the author did not use any AI-tools or -services. The author takes full responsibility for the content of the published article.

## Statement of copyrights

## Appendix I

In Gabiña et al[13] the apparent activity is given for a sphere of radius $r$ and a uniform activity concentration contained in the same sphere. For calculation of the influence of the background activity concentration the modified contrast recovery coefficient can be calculated by a similar calculation. The apparent activity is still calculated by integration over the inner sphere, while the convolution is calculated for the outer sphere including the cold wall by replacing $r$ with $r+t$ in the original equation. Equation (A1) in Gabiña et al[13] for the apparent activity $A_{app}$ is modified to

$$A_{\text{app}} = \int_0^r c \cdot 4\pi\rho^2 d\rho \ = 2\pi a \left\{ \int_0^r \rho^2 \operatorname{erf}\left(\frac{r+t-\rho}{\sigma\sqrt{2}}\right) d\rho + \int_0^r \rho^2 \operatorname{erf}\left(\frac{r+t+\rho}{\sigma\sqrt{2}}\right) d\rho \right.$$
$$\left. - \frac{2\sigma}{\sqrt{2\pi}} \left[ \int_0^r \rho\, \mathrm{e}^{-\frac{(r+t-\rho)^2}{2\sigma^2}} d\rho - \int_0^r \rho\, \mathrm{e}^{-\frac{(r+t+\rho)^2}{2\sigma^2}} d\rho \right] \right\}, \tag{I.1}$$

where $t$ is the thickness of the cold wall. Evaluation of equation (I.1) is analog to the derivation given in the Appendix of Gabiña et al[13]. The wall thickness dependent contrast recovery coefficient $CRC_t$ is consequently given by

$$CRC_t = \frac{A_{\text{app}}}{\frac{4}{3}\pi r^3 a}\,, \tag{I.2}$$

which equals equation (6) and the unmodified equation for $CRC$ for $t$=0. CRC in Gabiña et al[13] is given by

$$CRC = \operatorname{erf}\left(\frac{2r}{\sigma\sqrt{2}}\right) - \frac{1}{\sqrt{2\pi}}\frac{\sigma}{r}\left(3 - \mathrm{e}^{-\frac{2r^2}{\sigma^2}}\right) + \frac{1}{\sqrt{2\pi}}\left(\frac{\sigma}{r}\right)^3\left(1 - \mathrm{e}^{-\frac{2r^2}{\sigma^2}}\right). \tag{I.3}$$

***Table 1**: Values for the RC difference $\Delta RC$ expressed as $\delta CRC = SBR \cdot \Delta RC$ due to a wall thickness of 1 mm for spheres with a diameter d of 10, 37 and 60 mm. The three approximations from equation (5), (7) and (8) are also listed. The $\delta CRC_1$-approximation of equation (7) is only valid up to approximately the position of maximum, which is at FWHM/d≈0.74. Invalid values are shown in italic.*

| **FWHM (mm)** | $\delta CRC$ **from eq. (3) & (6)** | $\delta CRC_1$ **approx. from eq. (7)** | $\delta CRC_{max}$ **approx. from eq. (8)** | $\delta CRC_{1st\ term}$ **approx. from eq. (5)** |
|---|---|---|---|---|
| ***d*=10 mm** | | | | |
| **2** | 0.1660 | 0.2593 | 0.5534 | |
| **4** | 0.1917 | 0.2187 | 0.2767 | |
| **6** | 0.1737 | 0.1780 | 0.1845 | |
| **8** | 0.1436 | *0.1374* | 0.1383 | 0.3000 |
| **10** | 0.1121 | *0.0967* | 0.1107 | |
| **12** | 0.0852 | *0.0560* | 0.0922 | |
| **16** | 0.0489 | *-0.0253* | 0.0692 | |
| **20** | 0.0292 | *-0.1066* | 0.0553 | |
| ***d*=37 mm** | | | | |
| **2** | 0.0464 | 0.0781 | 0.5534 | |
| **4** | 0.0599 | 0.0751 | 0.2767 | |
| **6** | 0.0632 | 0.0722 | 0.1845 | |
| **8** | 0.0634 | 0.0692 | 0.1383 | 0.0810 |
| **10** | 0.0623 | 0.0662 | 0.1107 | |
| **12** | 0.0606 | 0.0633 | 0.0922 | |
| **16** | 0.0561 | 0.0573 | 0.0692 | |
| **20** | 0.0510 | 0.0514 | 0.0553 | |
| ***d*=60 mm** | | | | |
| **2** | 0.0287 | 0.0489 | 0.5534 | |
| **4** | 0.0376 | 0.0477 | 0.2767 | |
| **6** | 0.0403 | 0.0466 | 0.1845 | |
| **8** | 0.0411 | 0.0455 | 0.1383 | 0.0500 |
| **10** | 0.0411 | 0.0444 | 0.1107 | |
| **12** | 0.0408 | 0.0432 | 0.0922 | |
| **16** | 0.0395 | 0.0410 | 0.0692 | |
| **20** | 0.0378 | 0.0387 | 0.0553 | |

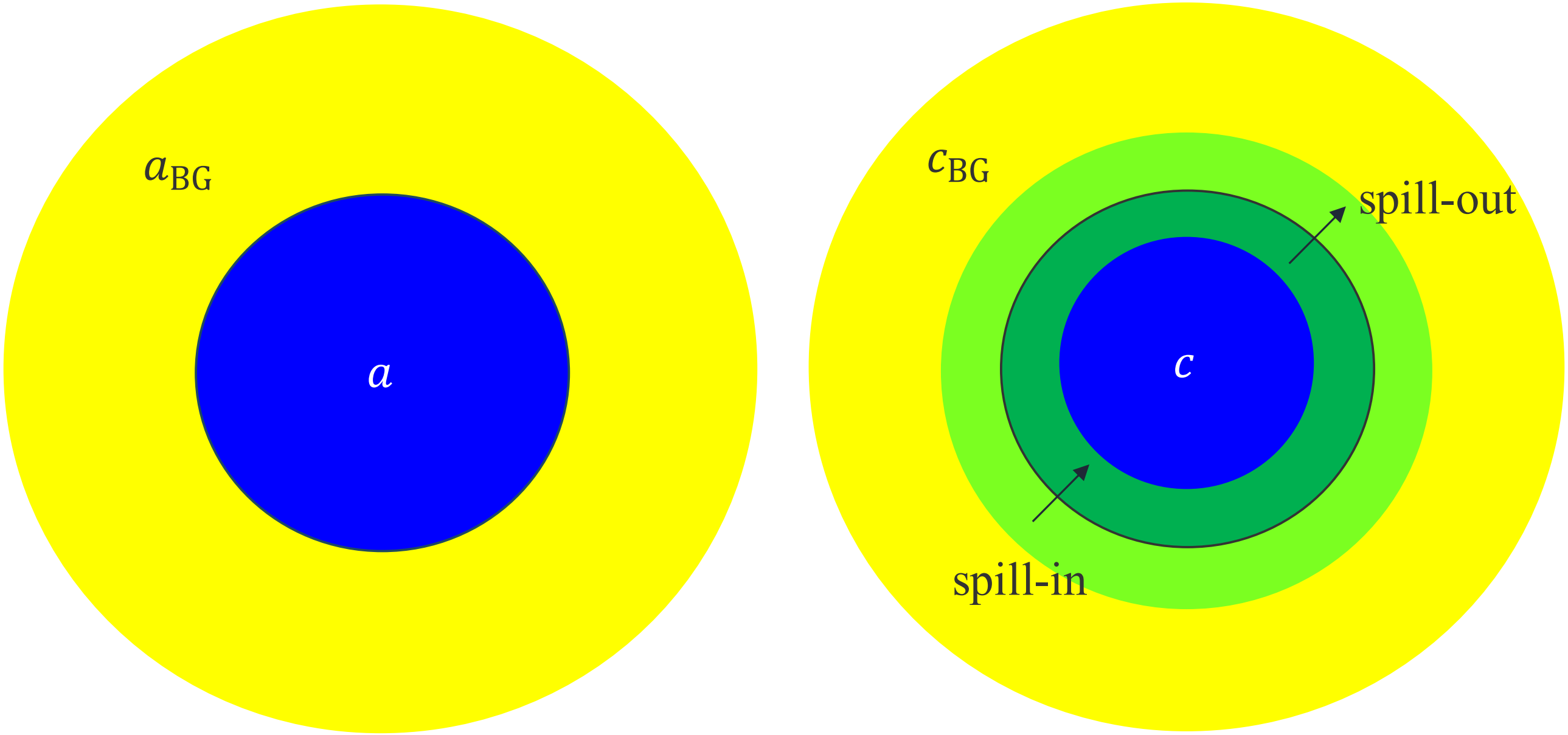

***Figure 1**: A (blue) disc (left) with true activity concentration a and a (yellow) background with true concentration $a_{BG}$. The imaged disc (right) and background illustrating the effect of limited spatial resolution. Measured concentrations are indicated by c for the disc and $c_{BG}$ for the background. In this case measured (far away) and true background concentrations are identical. Spill-in and spill-out is illustrated by the mixture of blue and yellow colors and shows a signal loss inside the disc.*

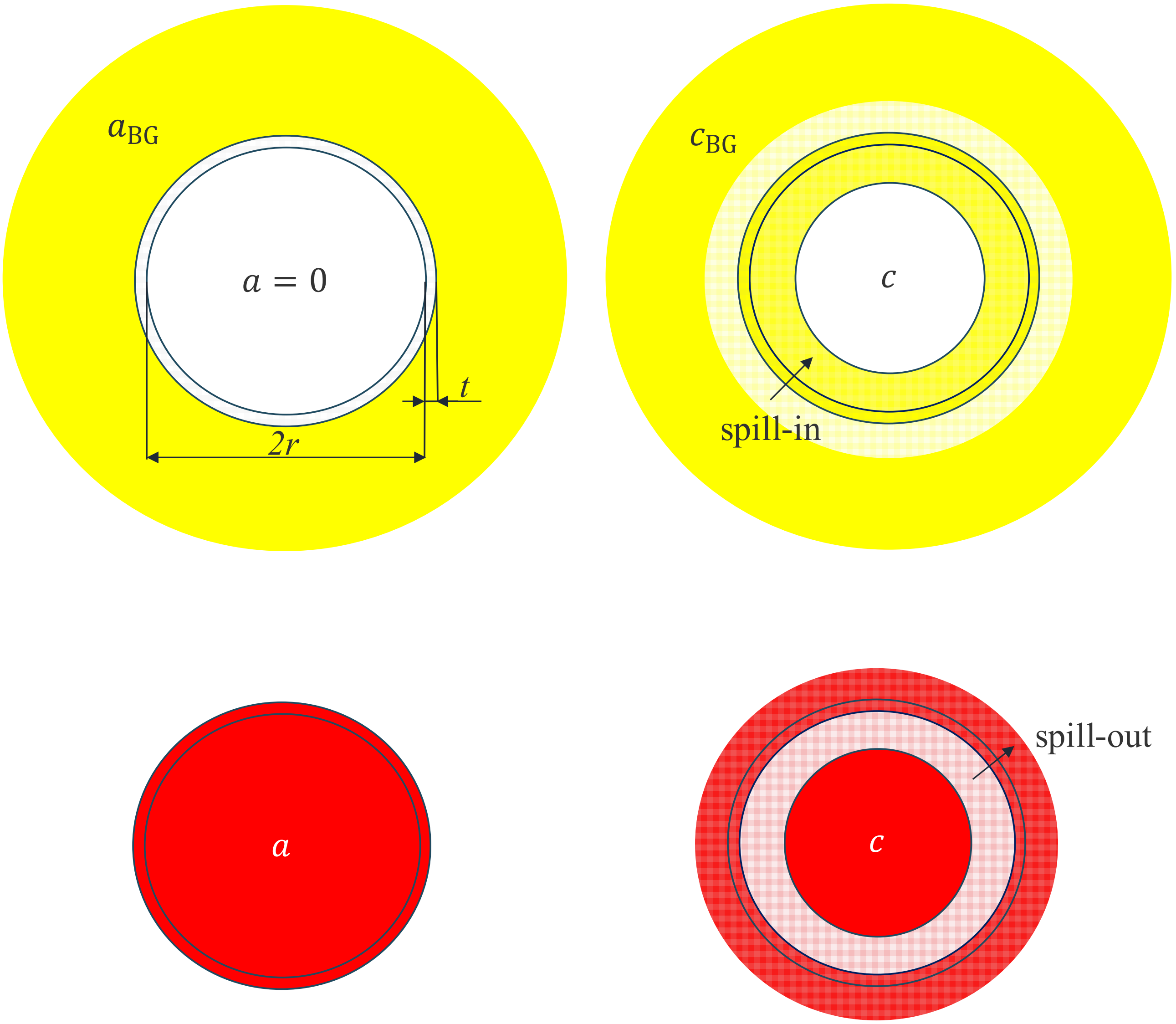

***Figure 2***: *Top row: A (white) disc (left) with zero true activity concentration a=0 and a (yellow) background with true concentration $a_{BG}$. The inner radius and wall thickness are denoted as r and t, respectively. The imaged disc (right) and background illustrating the effect of limited spatial resolution and the cold wall. Measured concentrations are indicated by c for the disc and $c_{BG}$ for the background. Spill-in is reduced by the presence of the cold wall. Bottom row: The same disc and background as in the top row with reversed activity concentrations for the calculation of $CRC_t$. This corresponds to the*

*"hot spot" RC situation with no background activity. However, the spill-out from the (inner) disc is reduced by the presence of the hot wall, since the activity (after reversing the concentrations from the top row) is contained in both the disc and wall.*

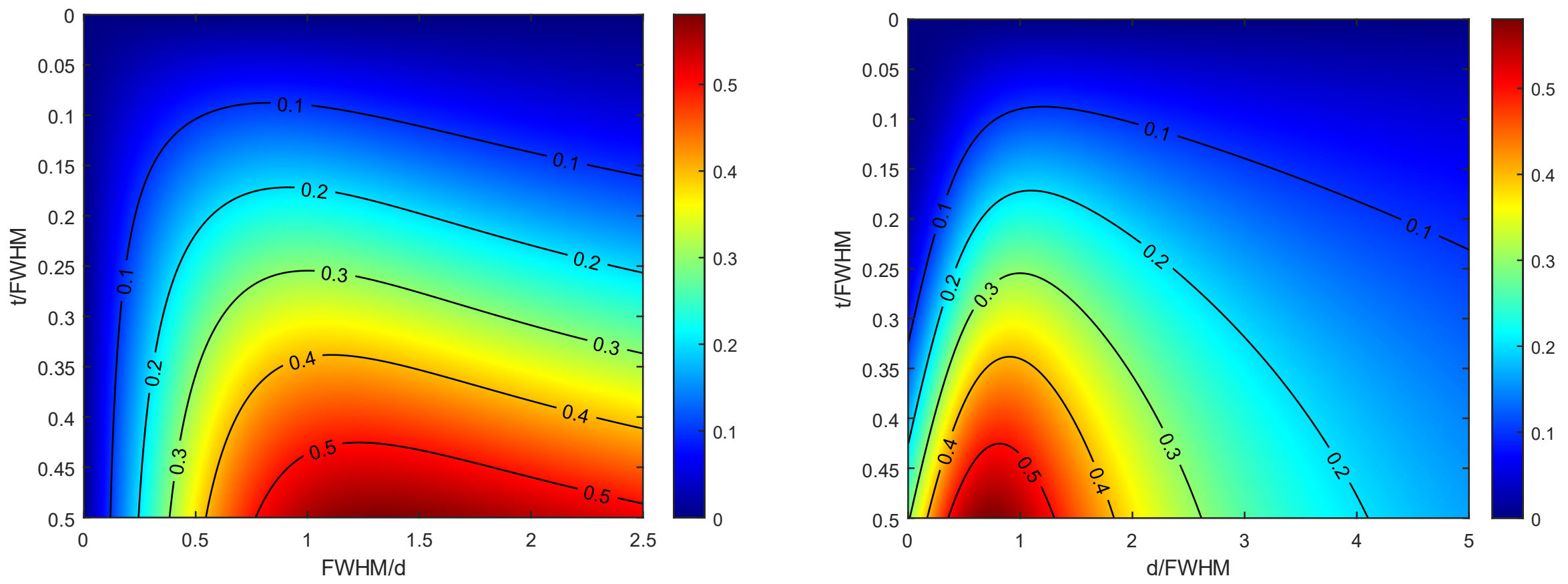

***Figure 3**: Values for the influence of the wall thickness on RC expressed as $\delta CRC = SBR \cdot \Delta RC$ in equations (3) and (6). Ratios of Full Width Half Maximum (FWHM) and diameter (d) are on the horizontal axes, while the ratios between wall thickness t and FWHM are on the horizontal axis. The change in RC is given by dividing the displayed δCRC-values with the actual SBR-value.*

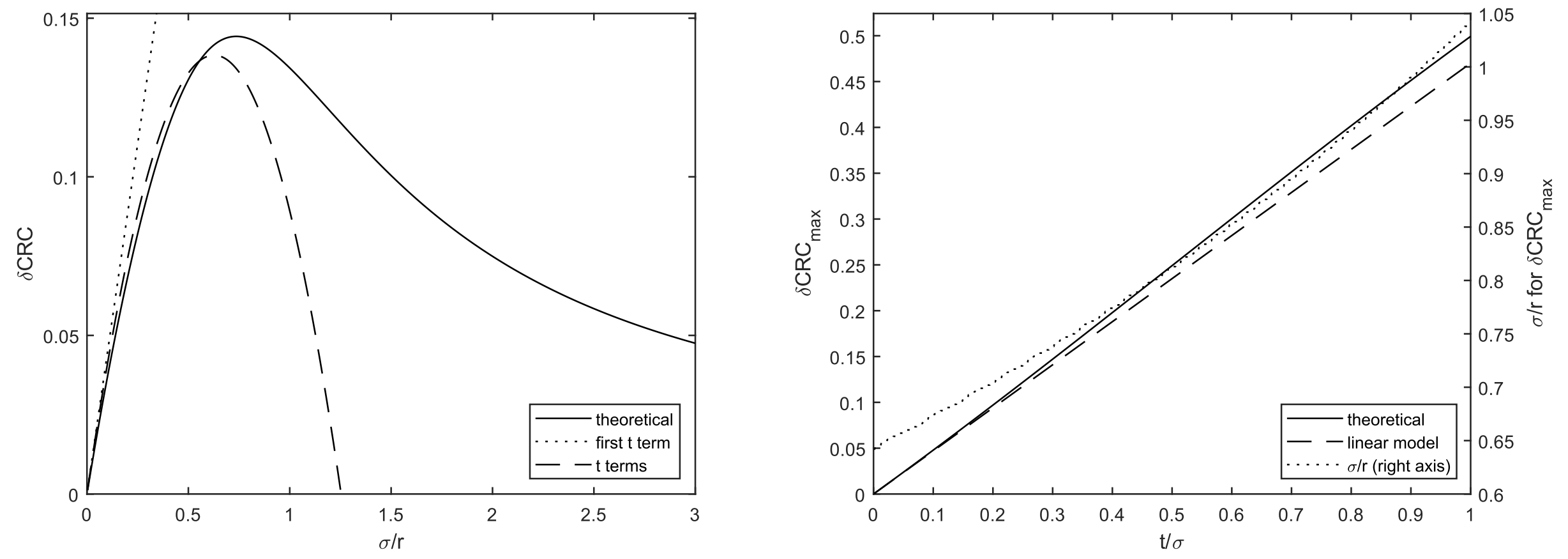

***Figure 4:*** *In left panel the influence of 1 mm cold walls is shown as δCRC for 8 mm FWHM as a function of the (reciprocal) radius. Three curves are shown. The "theoretical" curve is based on equation (3) and (6). The other two curves are based on equation (7), where "first term" refers to the first term in equation (7), which equals equation (5), and "t terms" to equation (7). In the right panel, the maximum "theoretical" value of δCRC and its "linear" approximation by equation (8) is shown. The position of the maximum is displayed according to the right vertical axis, where a value of 0.63 is expected for the linear approximation.*

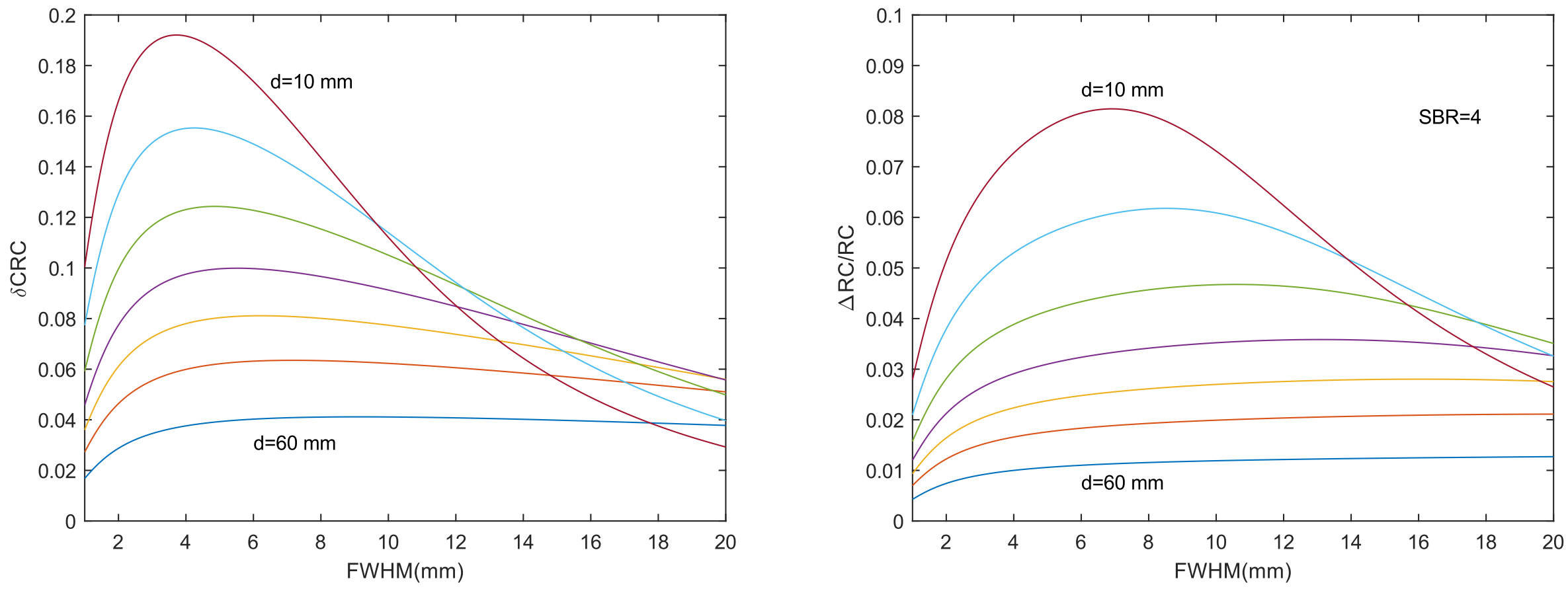

***Figure 5****: Changes in RC for the seven spheres with a diameter between 10 and 60 mm as function of FWHM for a wall thickness of 1 mm. In the left panel the absolute change ΔRC times SBR, which equals δCRC, is shown. The right panel shows the relative change in RC, denoted for SBR=4. The relative change ΔRC/RC equals* $SBR^{-1} \cdot \delta CRC/RC$*, and RC is given as* $RC_t$ *by equation (3). The different spheres can be identified by looking at FWHM-values between 2 and 8 mm, where the smallest sphere is on top and the largest sphere in the bottom. Sphere diameters are 10, 13, 17, 22, 28, 37 and 60 mm.*

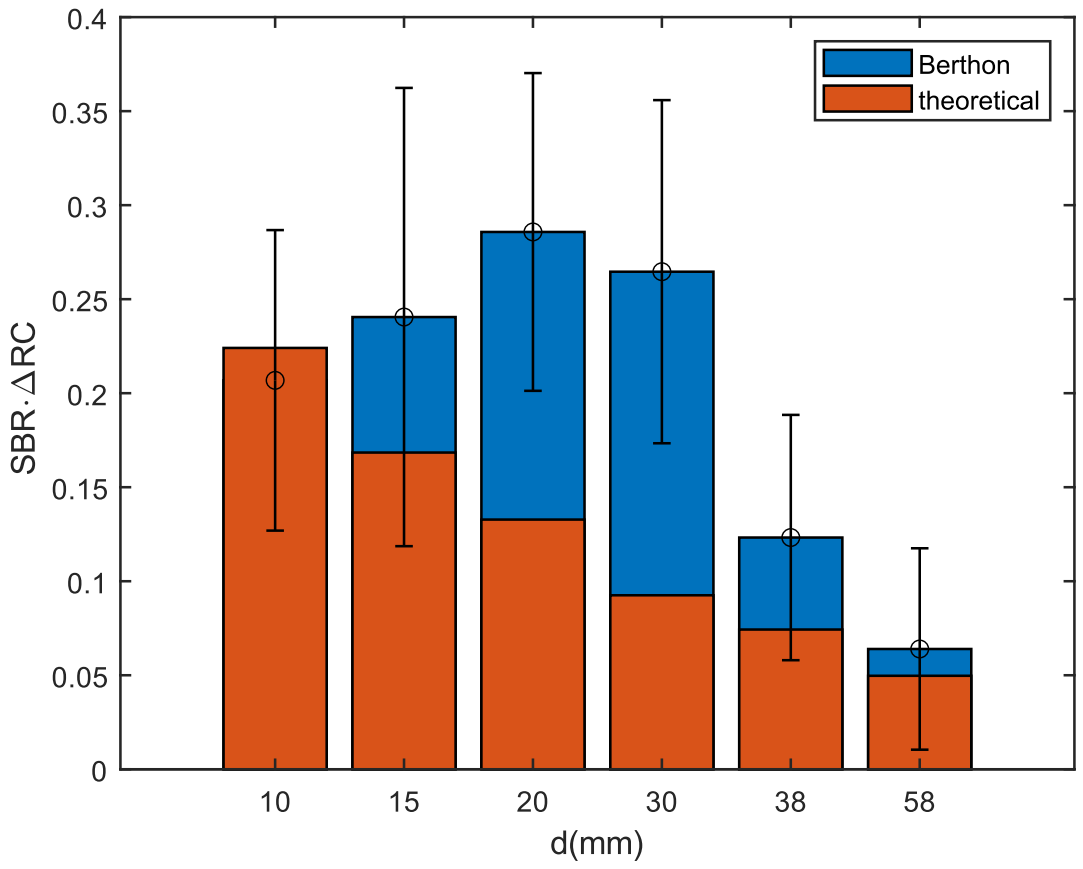

***Figure 6**: Difference in measured RC-values (for the mean) ΔRC between thin (~0.18 mm) and thick (~1.5 mm) walls, corrected for background by multiplying the measured difference with the SBR, which theoretically corresponds to ΔRC-values for SBR=1. Measurements[11] for six spheres with different diameters are shown as blue bars with error bars indicating 1.96 standard deviations of the mean. The orange bars show the theoretical values for SBR·ΔRC for FWHM=6 mm, which equal the difference between $CRC_t$ for thick and thin walls according to equation (3) and (6).*